\documentclass[fleqnp,10pt]{wlscirep}
\usepackage[utf8]{inputenc}
\usepackage[T1]{fontenc}
\usepackage{mathtools}
\usepackage{float}
\usepackage{amsmath}
\usepackage{comment}
\usepackage{bookmark}
\usepackage{xcolor}
\title{Discontinuous phase transitions in the $q$-voter model with generalized anticonformity on random graphs.}

\author[1]{Angelika Abramiuk-Szurlej}
\author[1]{Arkadiusz Lipiecki}
\author[1]{Jakub Paw{\l}owski}
\author[1,*]{Katarzyna Sznajd-Weron}
\affil[1]{Department of Theoretical Physics, Faculty of Fundamental Problems of Technology, Wrocław University of Science and Technology, 50-370 Wrocław, Poland}
\affil[*]{katarzyna.weron@pwr.edu.pl}


\begin{abstract}
We study the binary $q$-voter model with generalized anticonformity on random Erdős–Rényi graphs. In such a model, two types of social responses, conformity and anticonformity, occur with complementary probabilities and the size of the source of influence $q_c$ in case of conformity is independent from the size of the source of influence $q_a$ in case of anticonformity. For $q_c=q_a=q$ the model reduces to the original $q$-voter model with anticonformity.  Previously, such a generalized model was studied only on the complete graph, which corresponds to the mean-field approach. It was shown that it can display discontinuous phase transitions for $q_c \ge q_a + \Delta q$, where $\Delta q=4$ for $q_a \le 3$ and $\Delta q=3$ for $q_a>3$. In this paper, we pose the question if discontinuous phase transitions survive on random graphs with an average node degree $\langle k\rangle \le 150$ observed empirically in social networks.
Using the pair approximation, as well as Monte Carlo simulations, we show that discontinuous phase transitions indeed can survive, even for relatively small values of $\langle k\rangle$. Moreover, we show that for $q_a < q_c - 1$ pair approximation results overlap the Monte Carlo ones. On the other hand, for $q_a \ge q_c - 1$ pair approximation gives qualitatively wrong results indicating discontinuous phase transitions neither observed in the simulations nor within the mean-field approach. Finally, we report an intriguing result showing that the difference between the spinodals obtained within the pair approximation and the mean-field approach 
follows a power law with respect to $\langle k\rangle$, as long as the pair approximation indicates correctly the type of the phase transition.
\end{abstract}
\begin{document}

\flushbottom
\maketitle
%
%
\thispagestyle{empty}

\section{Introduction}
For researchers outside of statistical physics, it may seem strange why do physicists try so hard to distinguish between different types of phase transitions and why do they search so eagerly for discontinuous ones in the models of opinion dynamics \cite{Li:etal:16,Vie:Cro:16,Che:etal:17,Tuz:Fer:Equ:18,Enc:etal:18,Enc:etal:19,Abr:Paw:Szn:19,Chm:etal:20,Kra:20,Gra:Kra:20}. In the context of social systems, such efforts can be justified by the empirically confirmed existence of phenomena strictly related to discontinuous phase transitions, such as tipping points, critical mass, and hysteresis \cite{Sche:Wes:Bro:03,Bis:15,Pru:etal:18,Str:Liz:17,Cen:etal:18}. 

Several factors may facilitate discontinuous phase transitions, such as: (a) increasing the number of states that the system's entities (agents, particles, etc.) can take \cite{Wu:82,Li:etal:16,Oes:Pir:Cro:19,Now:Sto:Szn:21}, (b) introducing an aversion to change, which can be related to the size of the group needed to influence the agent \cite{Nyc:Szn:Cis:12,Per:etal:18} or another kind of inertia \cite{Che:etal:17,Enc:etal:18}, (c) higher dimensionality of the system \cite{Hen:Hin:Lub:08}, higher system degree  \cite{Enc:etal:18}, or the larger number of layers in the multiplex \cite{Chm:Szn:15}. Finally, the presence of the annealed instead of the quenched disorder usually promotes discontinuous phase transitions \cite{Aiz:Weh:89,Bor:Mar:Mun:13,Vil:Bon:Mun:14,Jed:17}.

Until recently, it seemed that in models of opinion dynamics, a type of nonconformity could also be a factor influencing the type of the phase transition. For example, in the case of the $q$-voter model \cite{Cas:Mun:Pas:09}, the presence of independence (noise) was needed to induce discontinuous phase transitions and the anticonformity itself was not sufficient \cite{Nyc:Cis:Szn:12,Jed:17,Jed:etal:20}. Analogous observation was done for the symmetrical threshold model \cite{Now:Szn:19}. However, two years ago it was shown that discontinuous phase transitions can occur also in the $q$-voter model with anticonformity if we assume that the size of the group of influence needed for conformity $q_c$ is independent of the size of the group of influence needed for anticonformity $q_a$\cite{Abr:Paw:Szn:19}.

Such a generalized model, which reduces to the basic $q$-voter model with anticonformity for $q_c=q_a=q$, was studied so far only on the complete graph. It was shown that in the case of the complete graph, which corresponds to the mean-field approach (MFA), it displays discontinuous phase transitions for $q_c \ge q_a + \Delta q$, where $\Delta q=4$ for $q_a \le 3$ and $\Delta q=3$ for $q_a>3$. However, real social networks, especially the friendship ones, do not have a complete graph's structure. It was predicted within the social brain hypothesis and then confirmed empirically that humans have an average of about $150$ relationships and this is true for both offline and online (Facebook, Twitter) networks \cite{Zho:etal:05,Dun:etal:2015,Car:Kas:Dun:16}. It means that the average degree $\langle k\rangle$ of a node in the social network is substantially smaller than the size $N$ of the entire social network, whereas for the complete graph $\langle k\rangle=N-1$. On the other hand,  the higher network degree facilitates discontinuous phase transitions \cite{Enc:etal:18}. Therefore, the natural question that arises here is if the model with generalized anticonformity can display discontinuous phase transitions for $\langle k\rangle<<N$. 

To answer the question, we study the model on random Erdős–Rényi (ER) graphs with an average node degree $\langle k\rangle$ that corresponds to those that were found empirically in social networks \cite{Dun:etal:2015,Car:Kas:Dun:16}. We decided on such a simple structure because in such a case the pair approximation (PA) should be more valid than for networks with a higher clustering coefficient \cite{Gle:13,Jed:17,Jed_war:20,Chm:Gra:Kra:18,Rad:San:20}. However, to validate the analytical results, we conduct also Monte Carlo simulations, because it was reported recently that PA can give invalid results even on ER graphs if the mean degree of nodes is small and comparable with the size of the influence group $q$ \cite{Jed:etal:20,Gra:Kra:20,Kra:21}. In such a case, the results can be wrong not only quantitatively, but even qualitatively, indicating discontinuous phase transitions, which are not observed in the computer simulations. 

\section{Model}
We consider a system of $N$ interconnected voters (also called agents or spins). Each of them is characterized by a single dynamical binary variable $s(x$,~$t) = j$, where $j=+1$~$(\uparrow)$ or $j = - 1$~$(\downarrow)$, $x=1$, $\ldots$, $N$, and $t$ denotes time. From the social point of view, $s(x$,~$t)$ represents an opinion of an agent placed at node $x$ at time $t$ on a given topic measured in the two-point psychometric scale (yes/no, agree/disagree). We use $j$ to describe the state of a single spin for consistency with earlier papers in which PA was used for the $q$-voter model \cite{Jed:17,Jed:Szn:20}.

Agents are placed in the vertices of an arbitrary graph and can interact only with those agents that are directly linked to them. Following \cite{Nai:Mac:Lev:00} we use here the term \textit{source} for the group of influence and the term \textit{target} for the voter, who is influenced (the active one). We are taking into account two types of social response, conformity and anticonformity, and we use the annealed (situation) approach \cite{Jed:Szn:17}. It means that each agent can behave in two distinct ways with complementary probabilities: it can copy the opinion of the source or take the opposite one.

The social influence in this model is imposed only by the unanimous group of $q$ agents that are randomly chosen (without repetitions) out of $k_x$ nearest neighbors of a target at node $x$. Theoretically, it may happen that a given voter at node $x$ cannot be influenced at all, because its degree $k_x<q$. However, in social reality and our study, such a situation appears rarely. The typical size of the influence group $q$ varies from a few to a dozen people \cite{Bon:05}. On the other hand, the real egocentric personal social networks consist on average of $150$ relationships of successively inclusive layers at $5, 15, 50$ and $150$ alters \cite{Dun:etal:2015,Car:Kas:Dun:16}. Therefore, here we will also consider $\langle k\rangle \le 150$, with the condition $\langle k\rangle>q$.

In the original model, the source of influence consists of $q$ agents for both types of social response \cite{Nyc:Szn:Cis:12,Jed:Szn:17,Jed:etal:20}. However, recently it was proposed that the size of the source needed for conformity $q_c$ may not be the same as the size $q_a$ needed for anticonformity \cite{Abr:Paw:Szn:19}. Such a generalization led to discontinuous phase transitions, which were not observed before within the $q$-voter model with anticonformity, neither on the complete graph \cite{Nyc:Cis:Szn:12,Jed:Szn:17}, nor on the random graph \cite{Jed:etal:20}. 

\begin{figure}[htb]
	\centering
	\includegraphics[width = 8cm]{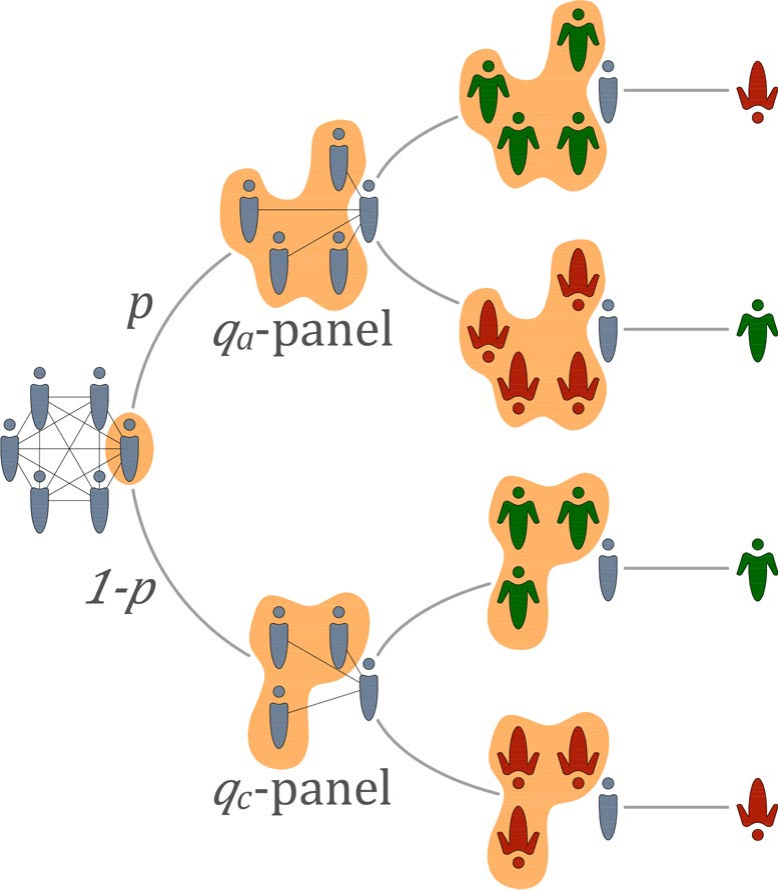}
	\caption{Schematic illustration of a single model's update. First, a target voter is chosen randomly and then with probability $p$ it acts as an anticonformist, whereas with probability $1-p$ as a conformist. In the first case (the upper branch in the plot), the source group of $q_a$ agents is chosen from the neighborhood ($q_a$-panel) of the target voter and if they are all in the same state then the active voter takes a state opposite to the $q_a$-panel. In the second case (the bottom branch in the plot), the source group of $q_c$ agents is chosen from the neighborhood ($q_c$-panel) of the target voter and if they are all in the same state then the active voter takes a state of the $q_c$-panel.} 
	\label{fig:model}
\end{figure}

The generalized model was studied so far only via the mean-field approach, which corresponds to the complete graph \cite{Abr:Paw:Szn:19}. Here, to answer the question posed in the Introduction, we study the model on random graphs with $\langle k\rangle \le 150$. As in all previous studies on the $q$-voter model, we use the random sequential updating (RSU) scheme, which is used to mimic the continuous time, and the algorithm of an elementary update is the following (see also Fig. \ref{fig:model}):
\begin{enumerate}
	\item a single vertex is chosen randomly $x \sim U\{1$,~$N\}$,
	\item a random number from the uniform distribution $r \sim U(0$,~$1)$ is drawn to decide about the type of response:
	\begin{enumerate}
		\item if $r<p$, where $p$ denotes probability of anticonformity, then a source of  $q_a$ agents is randomly drawn (without repetitions) out of $k_x$ target's direct neighbors; if the source is unanimous, the target takes the opposite state to the one of the source,
		\item otherwise, a source of  $q_c$ agents is randomly drawn (without repetitions) out of $k_x$ target's direct neighbors; if the source is unanimous, the target takes the same state as the source.
	\end{enumerate}
\end{enumerate}
As usually, a unit time, i.e., Monte Carlo step (MCS), consists of $N$ elementary updates.

\section{Pair Approximation}
In the previous paper, we considered the model on the complete graph, for which the MFA gives exact results \cite{Abr:Paw:Szn:19}. Here we consider the model on random ER graphs with the average node degree $\langle k\rangle<N-1$ and thus we use another analytical approach, so-called pair approximation \cite{Gle:13,Jed:17,Vie:etal:20}.

Within MFA, we assume that the global concentration of agents with positive opinion, defined as
\begin{equation}
c(t) \equiv c_\uparrow (t) =  \frac{N_{\uparrow}(t)}{N},
\end{equation}
approximates correctly the local one. Therefore, the probability to choose randomly an up-spin at time $t$ in the neighborhood of node $x$ is just equivalent to $c(t)$. However, this assumption is strictly valid only for the complete graph. A step further is to distinguish between the probability of choosing randomly an up-spin in the neighborhood of an up-spin and in the neighborhood of a down-spin. To make this distinction, the concentration of active bonds \(b(t)\) is introduced, with an active bond defined as a bond between opposite spins.

As shown in \cite{Jed:17,Jed:Szn:19} the set of equations that describes the evolution of the system within PA is universal for all single-spin flip binary dynamics:
\begin{equation}
\left\{\begin{aligned}
\frac{d c}{d t} &=\sum_{j\in\{\uparrow,\,\downarrow\}}c_j\sum_k P(k)\sum_{i=0}^{k}{k\choose i}\theta_j^i(1-\theta_j)^{k-i}f(\ldots)\Delta_c, \\
\frac{d b}{d t} &=\sum_{j\in\{\uparrow,\,\downarrow\}}c_j\sum_k P(k)\sum_{i=0}^{k}{k \choose i}\theta_j^i(1-\theta_j)^{k-i}f(\ldots)\Delta_b,
\end{aligned}\right.
\label{eq:system}
\end{equation}
where $k$ is a degree of a given node, $j$ is the state of this node, $i \in \{0$, $1$, $\ldots$, $k\}$ is the number of active edges attached to this node, $P(k)$ is the degree mass probability function, $\theta_j$ is the conditional probability of selecting a node that is in the opposite state to its neighbor in a state $j$, and $\Delta_c=-j$ and $\Delta_b=2(k-2i)/\langle k\rangle$ are rescaled elementary changes in corresponding quantities per one Monte Carlo step. {Moreover, for all single-flip binary dynamics $\theta_{\uparrow}=b/2c$ and $\theta_{\downarrow}=b/2(1-c)$ \cite{Jed:17,Jed:Szn:19}. Only the function $f(\ldots)$ depends on a given model and its parameters, since it describes the flipping probability of a node in a given state. Therefore, the only task in a broad class of binary opinion dynamics is to calculate $f(\ldots)$ \cite{Jed:Szn:19}.}

In our case, an opinion change happens only if $q_c$ neighbors with opposite opinions are drawn in case of conformity or $q_a$ neighbors with matching opinions are selected in case of anticonformity. Therefore, as the total number of neighbors with opposite opinions is equal to the number of active links $i$ of a given voter and the group of influence is chosen without repetition, the flipping probability takes the following form 
\begin{equation}
f(k,\,i,\,q_a,\,q_c,\,p) = (1-p)\frac{i!(k-q_c)!}{k!(i-q_c)!} + p\frac{(k-i)!(k-q_a)!}{k!(k-i-q_a)!}.
\label{eq:flip}
\end{equation}

Thus, in the limit of an infinite graph, the time evolution of this model is defined by a system of differential equations:
\begin{equation}
\left\{\begin{aligned}
\frac{d c}{d t} &= (1-p)\big[(1-c)\theta_\downarrow^{q_c} - c\theta_\uparrow^{q_c}\big] + p\big[ (1-c)(1-\theta_\downarrow)^{q_a} - c(1-\theta_\uparrow)^{q_a} \big],\\
\frac{d b}{d t} &= \frac{2}{\left<k\right>}\sum_{j\in\{\uparrow,\,\downarrow\}}c_j\left[(1-p)\theta_j^{q_c}\left(\left<k\right>-2q_c -2\left(\left<k\right>-q_c\right)\theta_j\right)
+ p\left(1-\theta_j\right)^{q_a}\left(\left<k\right>-2\left(\left<k\right>-q_a\right)\theta_j\right) \right].
\end{aligned}\right.
\label{eq:system_final}
\end{equation}

As always for the $q$-voter model, evolution equations (\ref{eq:system_final}) does not depend on the degree distribution $P(k)$ but only on the average degree $\langle k\rangle$ \cite{Jed:17,Jed:etal:20}. Hence, from now on we denote $\langle k\rangle$ by $k$ for brevity. Unfortunately, for $q_a \ne q_c$ we were not able to obtain the analytical solution of (\ref{eq:system_final}), even for the stationary state. Therefore, we solved it numerically. 

\section{Monte Carlo results}
To compare PA with MC results, we conducted simulations on sufficiently large graphs $N= 5 \cdot 10^5$. We used ensemble averaging, which means that for each set of parameters ($q_a$, $q_c$, $p$, $k$, $N$) we constructed $M$ independent graphs. It occurred that for a graph of size $N= 5 \cdot 10^5$ the number of samples $M=10$ was sufficient. As usual, the most problematic simulation parameter to choose was the "thermalization" time $\tau$ needed to reach the stationary state. It is well known that it increases dramatically near the critical point and therefore initially we used, similarly as for $N$ and $M$, different values of $\tau$ to check which one would be sufficient. Finally, we decided to use $\tau=5\cdot10^5$.

\begin{figure}[htb]
	\centering
	\includegraphics[width =\textwidth]{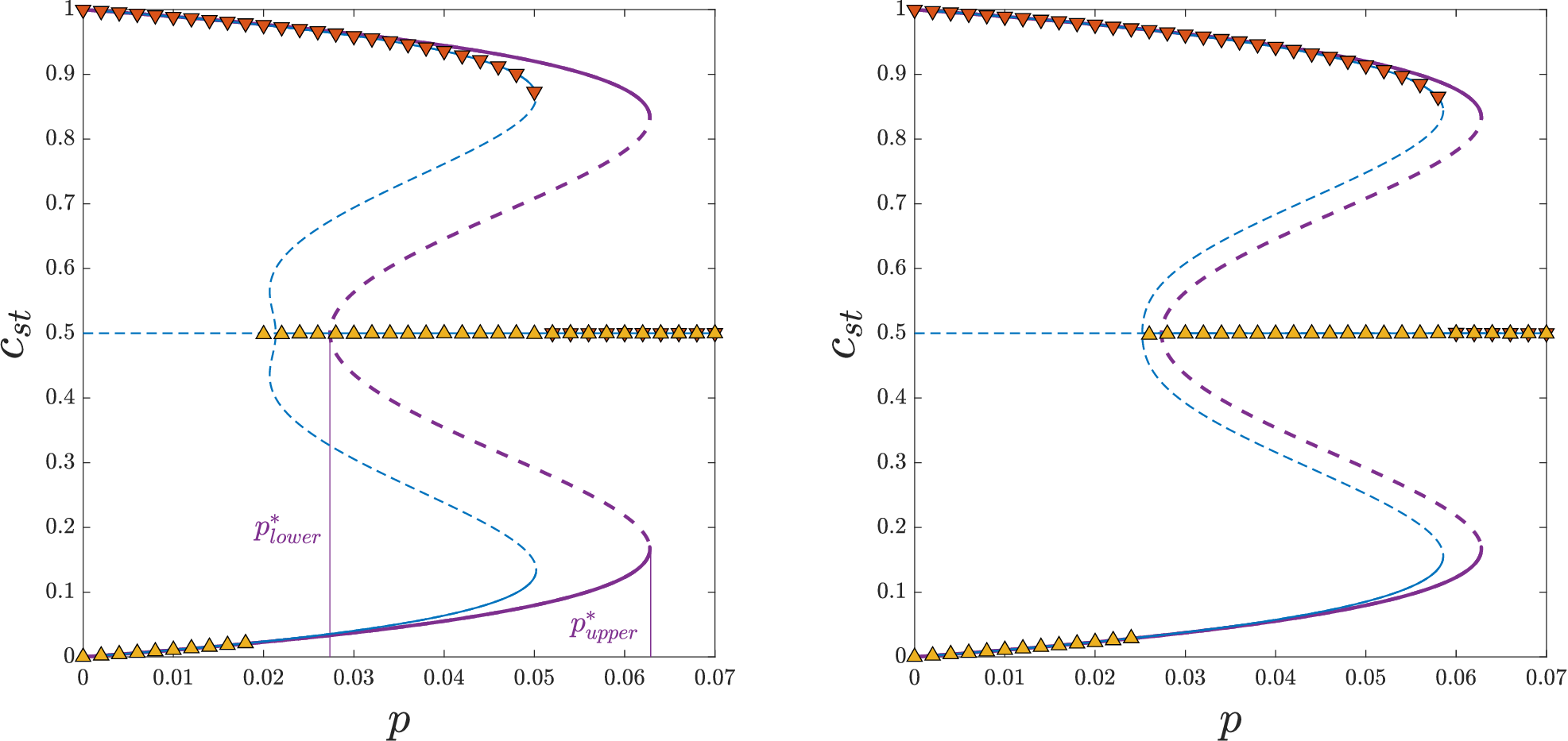}
	\caption{Dependence between the stationary concentration of up-spins $c_{st}$ and the probability of anticonformity $p$ obtained within MFA (thick purple lines), PA (thin blue lines) and MC simulations (symbols) for $q_a=4$ and $q_c=10$ and: $k=50$ (left panel), $k=150$ (right panel). Solid lines denote stable and dashed lines unstable fixed points. Results from the initial concentration of up-spins $c_0=1$ are presented in the upper parts of the panels (orange down-pointing triangles), whereas for $c_0=0.5$ in the bottom parts (yellow up-pointing triangles). {Thin purple vertical lines have been added to denote spinodals obtained from MFA. Error bars corresponding to the standard deviation of stationary concentration are smaller than the size of the markers.}} 
	\label{fig:C_p_qa4_qc10}
\end{figure}

The main goal of this paper is to check if the discontinuous phase transitions appear for realistic values of $k$. Hence, having in mind the condition $q_c, q_a <k$ and realistic values of $q$ and $k$ found in the empirical studies \cite{Bon:05,Dun:etal:2015,Car:Kas:Dun:16}, we use $k=50$, $150$ to show the dependence between the stationary value of the up-spins' concentration and the probability of anticonformity $p$. Because we want to check the type of the phase transition, we start from two different types of initial conditions, ordered $c_0=c(0)=1$ and disordered $c_0=c(0)=0.5$ ones. Such an approach allows us to see the hysteresis, which appearance is one of the indicators of a discontinuous phase transition. 

{As shown in Fig. \ref{fig:C_p_qa4_qc10}, starting from the ordered initial state $c_0=1$ the system remains in the ordered state for $p<p^*_{upper}$, i.e. below the upper spinodal,  and it reaches a disordered state for $p>p^*_{upper}$, i.e. above the upper spinodal.} 
On the other hand, for $c_0=0.5$ the system remains in the disordered state for $p>p^*_{lower}$, whereas for $p<p^*_{lower}$ it reaches one of two ordered states $c^{-}_{st}<0.5$ or $c^{+}_{st}=(1-c^{-}_{st})>0.5$,  equally likely. However, to avoid overlapping with the results for $c_0=1$, we plot both $c^{-}_{st}$ and $c^{+}_{st}=(1-c^{-}_{st})$ in the bottom part of the Figure. Both $p^*_{upper}$ and $p^*_{lower}$, depend on $k$, but we do not write it explicitly for brevity.

{The fact that $p^*_{lower} \ne p^*_{upper}$}, denotes the existence of the hysteresis: for $p \in (p^*_{lower}$,~$p^*_{upper})$ the stationary state depends on the initial one. Additionally, the jump of $c_{st}$, which can be easily converted to the order parameter $\phi=2c_{st}-1$, is seen. Both signatures of discontinuous phase transition, the hysteresis and the jump of the order parameter, are visible in PA and MC results. Moreover, as shown in Fig. \ref{fig:C_p_qa4_qc10}, MC overlaps PA results almost perfectly and approaches MFA with the increasing $k$, as expected. 

\begin{figure}[htb]
	\centering
	\includegraphics[width = \textwidth]{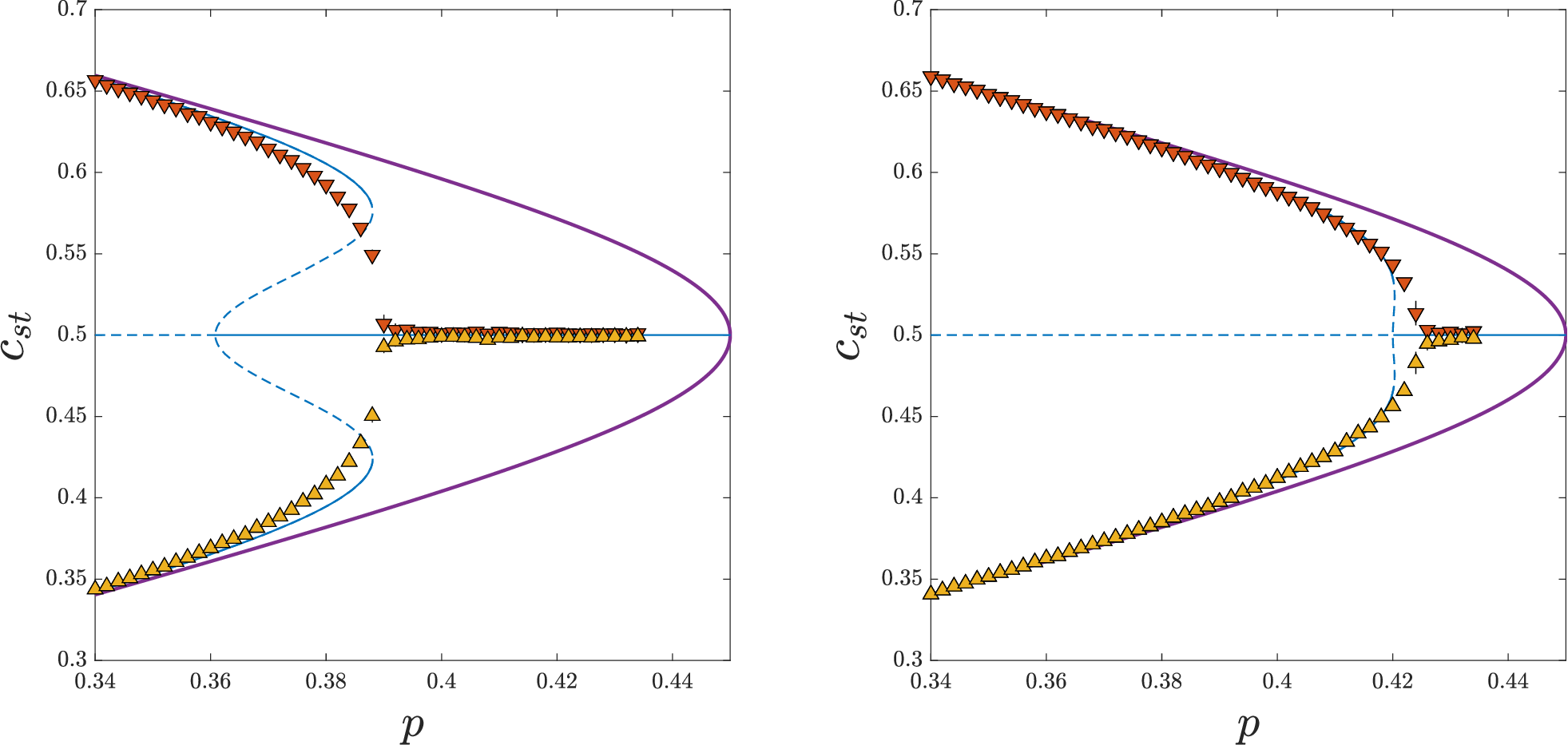}
	\caption{Dependence between the stationary concentration of up-spins $c_{st}$ and the probability of anticonformity $p$ obtained within MFA (thick purple lines), PA (thin blue lines) and MC simulations (symbols) for $q_a=q_c=q=10$ and : $k=50$ (left panel), $k=150$ (right panel). Solid lines denote stable and dashed lines unstable fixed points. Results from the initial concentration of up-spins $c_0=1$ are presented in the upper parts of the panels (orange down-pointing triangles), whereas for $c_0=0.5$ in the bottom parts (yellow up-pointing triangles). {Black vertical error bars represent the standard deviation of stationary concentration. They are visible only near the critical point and for all other values they are smaller than the size of the markers.}}
	\label{fig:C_p_qa10_qc10}
\end{figure}

\begin{figure}[htb]
	\hspace*{-0.5cm}
	\includegraphics[width =\linewidth]{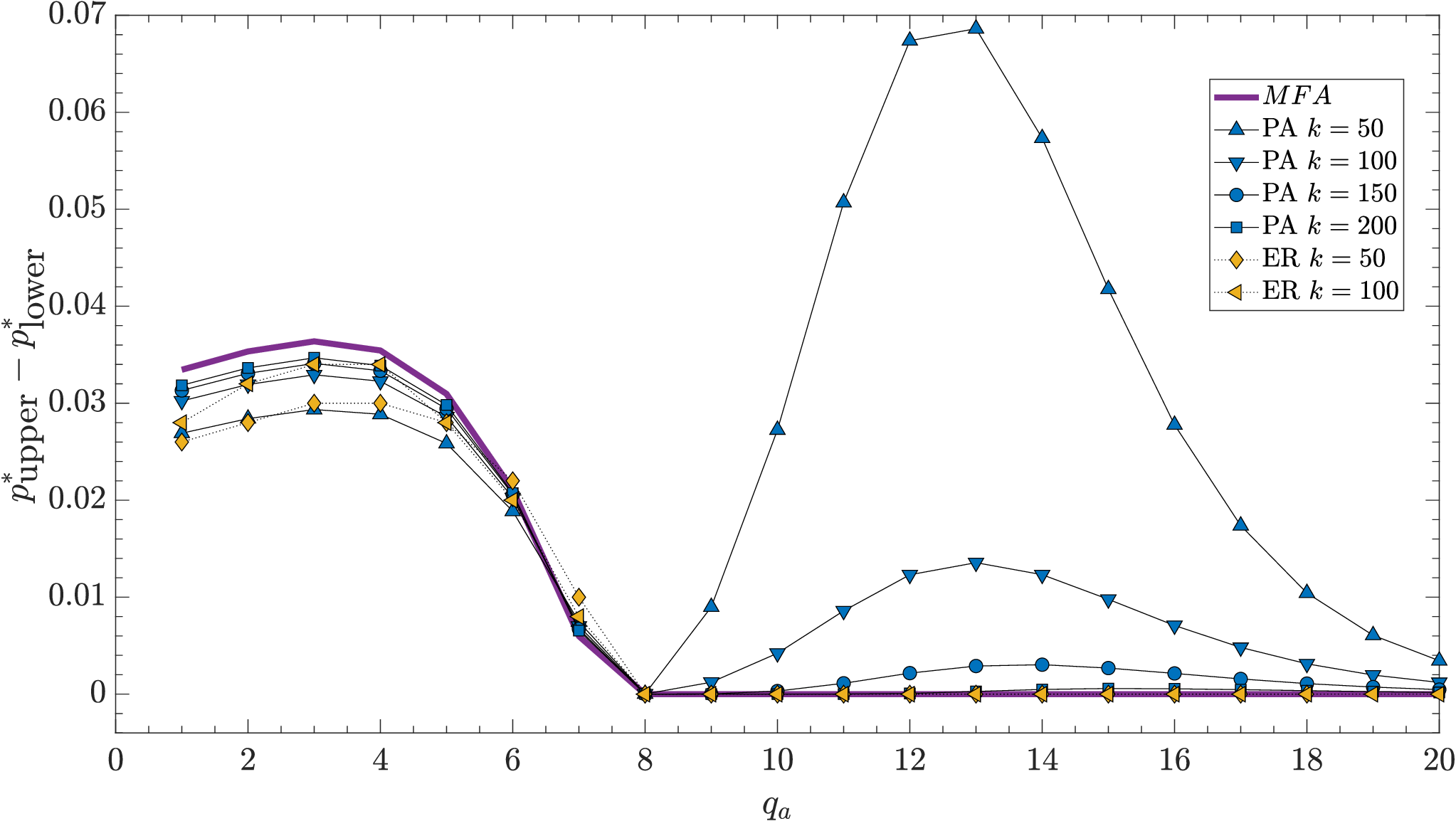}
	\caption{The width of the hysteresis, defined as the distance between the upper and the lower spinodal, as a function of \(q_a\) for \(q_c = 10\) within MFA, PA and MC simulations on ER graphs (indicated as ER in the legend). Thin black lines are just a guide to the eye.}
	\label{fig:hister_qa}
\end{figure}

However, it occurs that not for all values of parameters $q_a$, $q_c$ the agreement between PA and MC results is equally good, which is clearly seen in Fig. \ref{fig:C_p_qa10_qc10}. Within PA the hysteresis is seen, i.e., $p_{lower}^*<p_{upper}^*$, whereas within MC simulation no hysteresis is observed, i.e. $p_{lower}^*=p_{upper}^*$. As previously, PA results approach MFA ones with the increasing  $k$, and the hysteresis vanishes. To evaluate the values of the parameters for which PA gives correct predictions, we measure the width of the hysteresis, defined here as the distance between the upper and the lower spinodal $p_{upper}^*-p_{lower}^*$, as a function of $q_a$ for the fixed value of $q_c$ within all three methods: PA, MFA and MC. In Fig. \ref{fig:hister_qa} we see results for $q_c=10$ but we have checked also other values of $q_c$, which allows to draw general conclusions.

Let us recall first the results obtained analytically within MFA \cite{Abr:Paw:Szn:19}: a hysteresis indicating a discontinuous phase transition has been found for  $q_a \le q_c-3$. Indeed, we see in Fig. \ref{fig:hister_qa} that for $q_c=10$ and $q_a \le 7$, the hysteresis appears within MFA, PA, and MC. Moreover, for $q_a<q_c-1$ MC overlap PA results if only $k$ is sufficiently large (as in many other studies PA fails if the size of the influence group approaches $k$ \cite{Jed:17,Jed:etal:20,Gra:Kra:20,Kra:20}). 

\begin{figure}[htb]
	\centering
	\includegraphics[width = \textwidth]{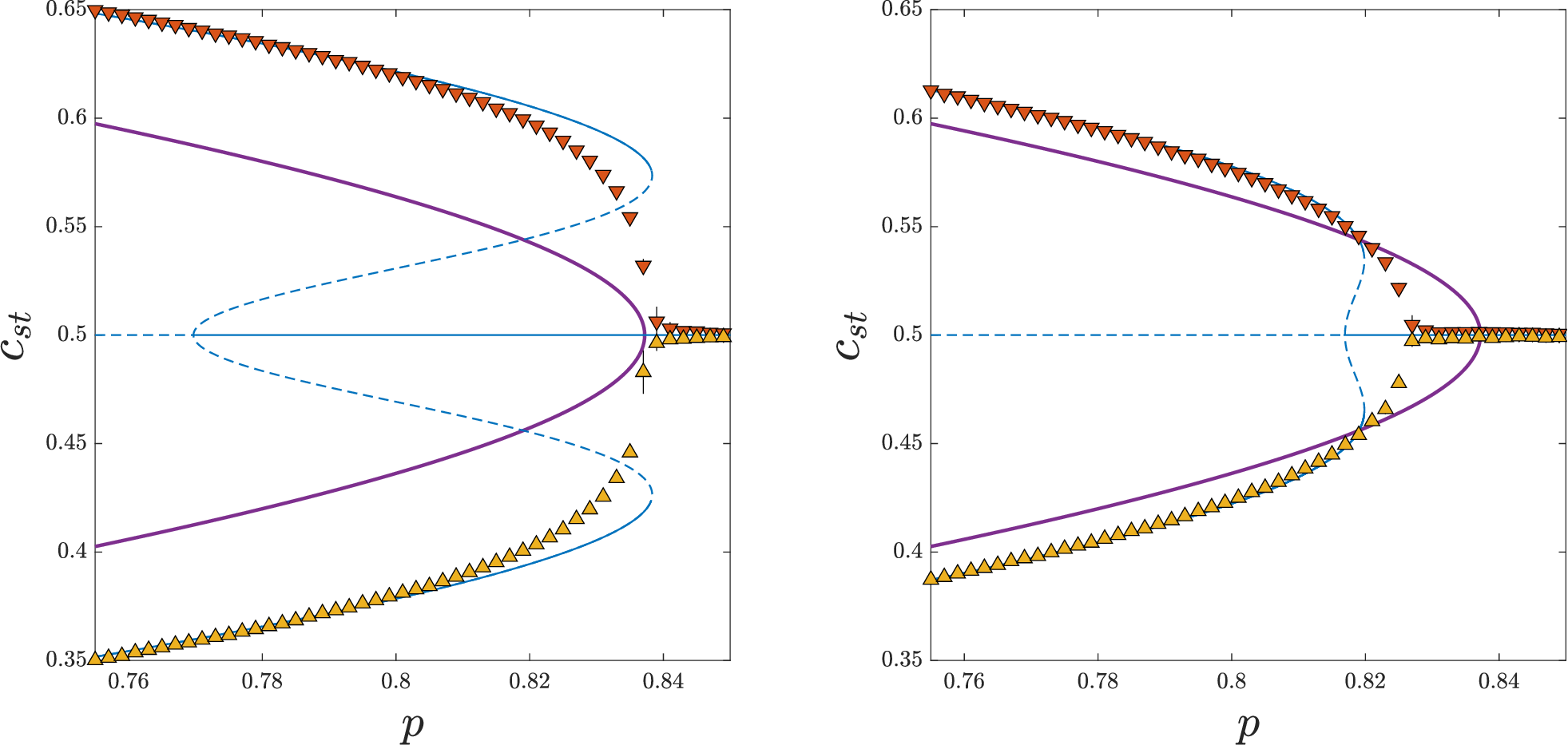}
	\caption{Dependence between the stationary concentration of up-spins $c_{st}$ and the probability of anticonformity $p$ obtained within MFA (thick purple lines), PA (thin blue lines) and MC simulations (symbols) for $q_a=13$, $q_c=10$ and : $k=50$ (left panel), $k=150$ (right panel). Solid lines denote stable and dashed lines unstable fixed points. Results from the initial concentration of up-spins $c_0=1$ are presented in the upper parts of the panels (orange down-pointing triangles), whereas for $c_0=0.5$ in the bottom parts (yellow up-pointing triangles). {Black vertical error bars represent the standard deviation of stationary concentration. They are visible only near the critical point and for all other values they are smaller than the size of the markers.}}
	\label{fig:C_p_qa13_qc10}
\end{figure}

For $q_a \ge q_c-1$ the hysteresis starts to increase within PA, then it reaches the maximum at a certain value $q_a^*=q_a^*(q_c)$, for example $q_a^*(10)=13$, and then it decreases. According to MFA, no hysteresis should be seen for  $q_a > q_c-3$ and indeed no hysteresis is seen for these values of parameters within MC. Even at $q_a^*=q_a^*(q_c)$, in which PA predicts the maximum width of the hysteresis, it does not occur in Monte Carlo simulations, which is clearly seen in Fig. \ref{fig:C_p_qa13_qc10}.  

\begin{figure}[htb]
	\centering
	\includegraphics[width=\textwidth]{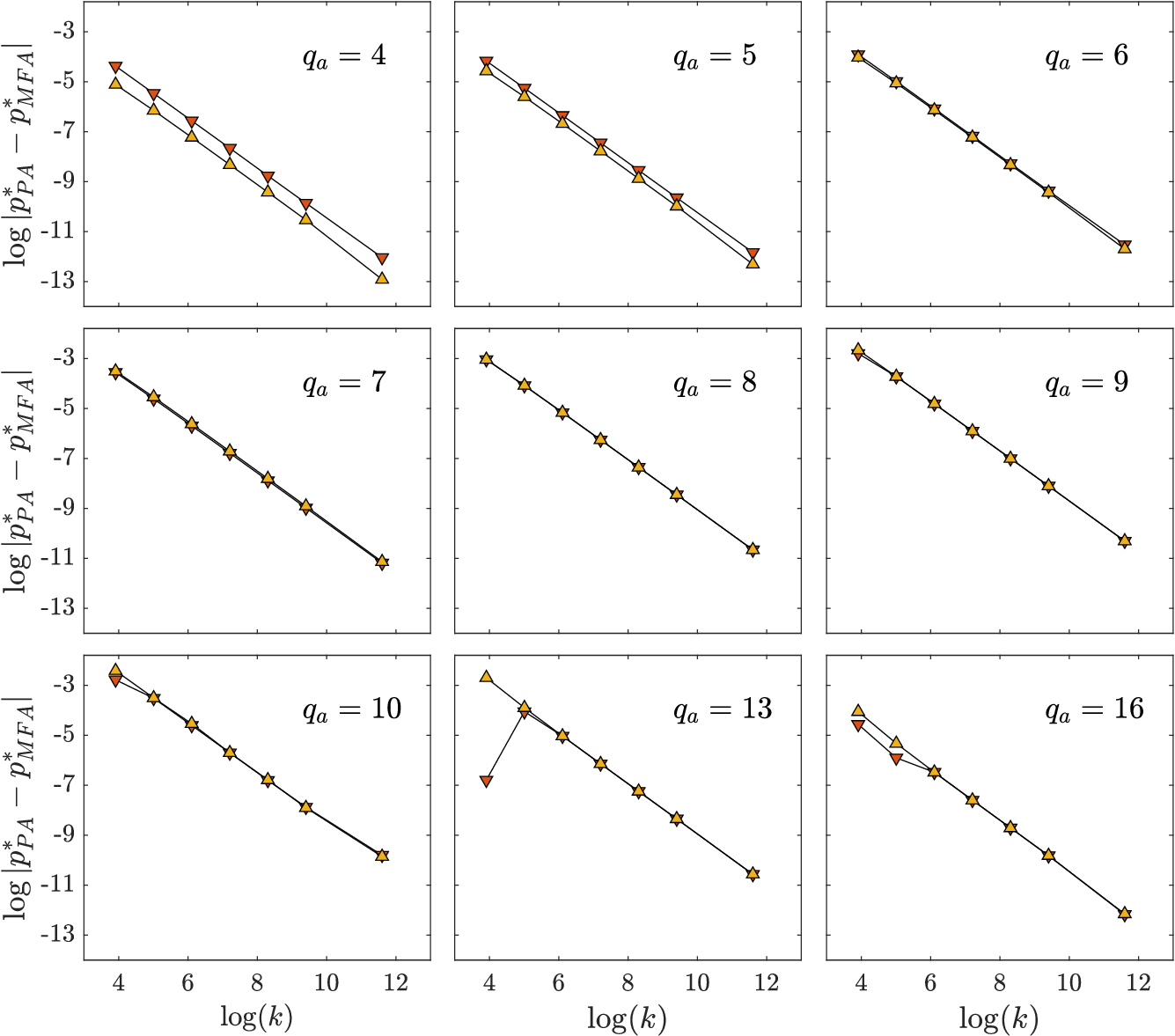}
	\caption{The absolute difference between {spinodals} obtained within PA and MFA as a function of $k=50$, $150$, $450$, $1350$, $4050$, $12150$, $109350$, for $q_c=10$ and different values of $q_a$ indicated in the upper right corners of the panels. Results for the upper spinodal $p_{upper}^*$ are represented by the orange down-pointing triangles, whereas for the lower one $p_{lower}^*$ by the yellow up-pointing triangles. Results are presented in the log-log scale (the natural logarithm is used here). {For the lower spinodal, the power-law appears in the whole examined range of $k$. On the other hand, the upper spinodal deviates from the power-law for $q_a \ge q_c-1$ for small values of $k$. It should be noticed that simultaneously for these values of parameters PA wrongly predicts discontinuous phase transition, i.e.  $p^*_{upper} \ne p^*_{lower}$, whereas MFA indicates the continuous one, i.e. $p^*_{upper} = p^*_{lower}$.}}
	\label{fig:powerlaw}
\end{figure}

In all presented results on Figs. \ref{fig:C_p_qa4_qc10} -- \ref{fig:C_p_qa13_qc10}, it is seen that the difference between PA and MFA results decreases with increasing $k$, which is an expected result. However, it is not that obvious how it decreases. Therefore, we decided to check how the {spinodals} $p_{lower}^*$, $p_{upper}^*$ obtained within PA deviates from the mean-field's ones. Results for $q_c=10$ are shown in Fig. \ref{fig:powerlaw}. First of all, we see that for both {spinodals} the power-law decay with $k$ is observed in case of $q_a<9$ (panels in two upper rows in Fig. \ref{fig:powerlaw}). We checked that the same phenomenon appears for other values of $q_c$. As long as $q_a<q_c-1$ PA predicts correctly the type of phase transition and simultaneously the power-law decay is observed. The small deviation from the power-law is seen only for small values of $k$, for which PA also fails. {Obviously, we cannot prove that the power-law holds for arbitrarily large $k$, but it exists for all examined values up till $k=109350$, which is a huge number from the social point of view.} 

For $q_a \ge q_c-1$ PA predicts discontinuous phase transitions, which are neither predicted by the MFA nor seen in the simulations, and simultaneously the power-law breaks down. It is probably worth to recall that the condition $q_a = q_c-1$ corresponds to the situation in which the same number of unanimous agents have to be selected to change the state of the system for both types of social response: 
\begin{itemize}
\item in case of anticonformity, the state of the system changes if a voter and the source of size $q_a$ are in the same state, so in total $q_a+1$ agents need to be unanimous,
\item in case of conformity, the state of the system changes  if a voter has an opposite state to  the source of size $q_c$, so in total $q_c$ agents need to be unanimous.
\end{itemize}
Unfortunately, the above remark neither explains why PA fails nor why the power-law breaks down. We decided to report here these intriguing results although unfortunately we are not able to explain them. We never saw this kind of analysis before, so we are not able to compare our results with others to gain some additional intuition. However, we hope that maybe some readers will be able to explain this intriguing phenomenon. 

\section{Summary}
In this paper, we investigated the $q$-voter model with generalized anticonformity on random graphs via Monte Carlo simulations, as well as pair approximation. It occurs that, similarly as within MFA, discontinuous phase transitions appear only if the size of the influence group $q_c$ needed for conformity is sufficiently larger than the size of the influence group $q_a$ needed for anticonformity, precisely for $q_a \le q_c-3$. For these values of parameters PA results overlap MC ones. 

Moreover, PA gives results in agreement with Monte Carlo simulations in the case of $q_a < q_c-1$ for any value of $q_a$. It predicts properly both the type of phase transition, as well as the values of the {spinodals}, which are overestimated by MFA for $k<<N$. However, for  $q_a \ge q_c-1$ PA wrongly predicts the type of phase transition and shows an unexpected non-monotonic behavior of the size of hysteresis. As expected,
PA results approach MFA ones with the increasing average node degree $k$. However, the difference between the results obtained within PA and MFA decreases with $k$ as power-law.

Going back to the initial aim of this paper -- we answered positively the question about the discontinous phase transions. Indeed, they can appear for some values of the model's parameters for a realistic average degree $k$. However, it is not the main take-home message we would like to pass to the reader because the model we consider here is just a toy model, one of many models of opinion dynamics \cite{Gal:etal:21}. In our opinion, the main take-home message of this paper is that PA calculations should always be treated with caution and it is advised to supplement it with simulation results.

\bibliography{Abramiuk_etal}

\section*{Acknowledgments}
This work has been partially supported by the National Science Center (NCN, Poland) through grants no. 2016/21/B/HS6/01256 and 2019/35/B/HS6/02530, and by Polish Ministry of Science and Higher Education through project “Diamentowy Grant” no.~DI2019 0150 49.  More extensive computations have been conducted using the PLGrid Infrastructure.

\section*{Author contribution statement}
A.A.-S. conducted analytical calculations within the pair approximation, A.L. conducted extensive Monte Carlo simulations, J.P. solved numerically analytical equations, K.Sz-W. designed and supervised the research. All authors reviewed and edited the manuscript.

\section*{Competing interests}
The authors declare no competing interests.

\end{document}